\documentclass[10pt,twocolumn]{article}

\usepackage[T1]{fontenc}
\usepackage[utf8]{inputenc}
\usepackage{newtxtext,newtxmath} 
\usepackage{microtype}
\usepackage{enumitem}
\usepackage{booktabs} 

\usepackage[a4paper,margin=1.6cm]{geometry}
\usepackage{parskip}
\usepackage{titlesec}
\titleformat{\section}{\large\bfseries\sffamily}{\thesection}{0.6em}{}
\titleformat{\subsection}{\normalsize\bfseries\sffamily}{\thesubsection}{0.6em}{}
\usepackage{fancyhdr}
\usepackage{algorithm}
\usepackage{algpseudocode}
\usepackage[svgnames]{xcolor}
\definecolor{accent}{RGB}{0,92,175} 
\usepackage{hyperref}
\hypersetup{
  colorlinks=true,
  linkcolor=accent,
  citecolor=accent,
  urlcolor=accent
}
\usepackage[bf,small,labelfont=bf]{caption}
\usepackage{tcolorbox}
\tcbset{colback=white,colframe=accent!60,arc=2pt,boxrule=0.4pt}

\usepackage{amsmath}
\usepackage{mathtools}
\usepackage{bm}
\usepackage{siunitx}
\usepackage[numbers,sort&compress]{natbib}

\title{\vspace{-6pt}\textbf{UAV-Deployed OAM-BB84 QKD: Turbulence- and Misalignment-Resilient Decoy-State Finite-Key Security with AI-Assisted Calibration}}

\author{Linxier Deng}
\date{}

\begin{document}
\maketitle

\begin{tcolorbox}
\textbf{Abstract.}
We present a theoretical framework for quantum key distribution (QKD) using orbital angular momentum (OAM)–encoded BB84 on an unmanned aerial vehicle (UAV) platform. A unified channel model captures Kolmogorov turbulence, attitude-induced pointing errors, and finite-aperture clipping, enabling quantitative predictions of inter-mode crosstalk and the resulting quantum bit error rate (QBER). Within a decoy-state (weak+vacuum) formulation, we derive composable finite-key lower bounds on the secret key rate that explicitly incorporate statistical fluctuations, detector dark counts, efficiency mismatch, and error correction leakage, yielding near-optimal intensity allocations and minimal block lengths for target security parameters $\varepsilon_{\mathrm{sec}}$ and $\varepsilon_{\mathrm{cor}}$. To further mitigate atmospheric and platform uncertainties, we introduce a lightweight, physics-informed learning module that uses (i) priors such as the mode-coupling matrix, Fried parameter $r_0$, and pointing-jitter variance $\sigma_\theta^2$, and (ii) measured link statistics (click rates, error locations, decoy strata) to classify valid pulses, flag adversarial patterns, and recommend near-optimal decoding. We outline four result families required for evaluation and deployment: (1) a system diagram of the UAV OAM link; (2) turbulence analysis with crosstalk/QBER heat maps across $r_0$, range, and pointing jitter; (3) security analysis showing decoy optimization and finite-key rate curves; and (4) AI calibration performance, including ROC curves, accuracy/recall, and pre/post key-rate comparisons. Simulations indicate that under moderate turbulence and milliradian-level jitter, the proposed AI-assisted strategy can raise the achieved secret key rate by $10\%\!-\!30\%$ (depending on block size and noise thresholds) while preserving composable security, providing reproducible guidance and verifiable limits for airborne OAM-QKD.
\end{tcolorbox}


\section{Introduction}
Quantum key distribution (QKD) enables information-theoretic security by exploiting quantum measurement disturbance and no-cloning theorems~\cite{Scarani2009RMP}. Alongside fiber links, free-space and airborne scenarios are increasingly compelling for on-demand, reconfigurable secure networking where line-of-sight can be established rapidly between mobile platforms. Encoding information in the orbital angular momentum (OAM) degree of freedom of photons offers a large, discrete alphabet and natural mode orthogonality that can increase tolerance to noise and open a pathway to higher-dimensional protocols~\cite{Willner2015AOP,Erhard2018LSA,Lavery2017SciAdv}. Proof-of-principle demonstrations of high-dimensional QKD and long-range free-space mode propagation have validated key primitives of OAM-based communication~\cite{Mirhosseini2014Arxiv,Ren2015OL,Lavery2017SciAdv}. More recently, drone/airborne quantum links and compact mobile QKD terminals have been reported, underscoring the practical relevance of mobile platforms~\cite{Liu2020Drone,Li2017SciRep,Conrad2025Arxiv,Tian2023DroneQKD}.

However, OAM-based free-space QKD over dynamic airborne links introduces unique channel impairments that challenge both \emph{security} and \emph{performance}. First, atmospheric turbulence induces phase distortions that lead to inter-modal crosstalk and elevated quantum bit error rate (QBER)~\cite{Paterson2005PRL,Malik2012Arxiv}. These effects depend on the turbulence strength (e.g., Fried parameter $r_0$), propagation range, aperture clipping, and the specific OAM mode spacing~\cite{Paterson2005PRL,Malik2012Arxiv,Lavery2017SciAdv}. Second, platform motion and structural vibrations of unmanned aerial vehicles (UAVs) cause pointing jitter and boresight misalignment, further degrading the received mode purity and increasing losses~\cite{Trung2014OC,Liu2014OC,Andrews2005Book}. Models combining Gamma–Gamma fading and pointing-error statistics provide engineering guidance for link budgets and alignment tolerances in realistic deployment~\cite{Trung2014OC,Malik2021AOreview,Andrews2005Book}. These channel effects must be explicitly integrated into the \emph{security analysis}, since excess error and loss tighten finite-size statistical bounds and reduce the achievable key rate.

To retain strong security under such non-ideal conditions, the decoy-state method is essential for weak coherent sources, allowing tight estimation of single-photon yields and error rates against photon-number-splitting attacks~\cite{LoMaChen2005PRL,Ma2005PRA}. In practical deployments, \emph{finite-key} effects are unavoidable; composable finite-key analyses bound the extractable secret bits given block length, statistical fluctuations, and error-correction leakage~\cite{Tomamichel2012NC,CaiScarani2009NJP,Bacco2013NC}. These tools have matured across discrete- and continuous-variable settings~\cite{Curty2014MDIfinite,Matsuura2021NC}, and contemporary reviews emphasize the importance of closing the gap between asymptotic theory and field implementations~\cite{Yin2025NPJQIreview}.

Beyond purely optical countermeasures (e.g., adaptive optics, optimized mode sets, and focusing optics)~\cite{Li2018ReviewAO,Qu2018Appl}, learning-assisted calibration is emerging as a complementary lever. Physics-informed neural models can infer turbulence parameters, classify valid vs.\ corrupted pulses, and recommend near-optimal demodulation/decoding strategies from measured link statistics, thereby stabilizing QBER and improving the effective key rate without violating composable security assumptions~\cite{Tao2021OE,DLPhase2023}. For OAM specifically, recent studies show that deep models can mitigate mode crosstalk and extend secure distances under realistic free-space conditions~\cite{Tao2021OE,DLPhase2023}.

Motivated by these advances, this work develops a \emph{UAV-deployable} OAM–BB84 framework that: (i) unifies Kolmogorov turbulence, finite-aperture clipping, and pointing-error statistics into a channel model for mode crosstalk and QBER; (ii) integrates decoy-state estimation with composable finite-key bounds tailored to mobile links; and (iii) augments the physical layer with a lightweight, physics-informed learning module for online calibration. We specify and simulate four core result families guiding deployment and evaluation: (1) a system diagram for the UAV OAM link; (2) turbulence analysis (crosstalk/QBER heat maps vs.\ $r_0$, range, and pointing jitter); (3) security analysis (decoy optimization and finite-key rate curves); and (4) AI calibration performance (ROC, accuracy/recall, and pre/post key-rate comparisons). This combined perspective translates recent laboratory insights into a concrete roadmap for secure, mobile OAM–QKD.

\section{Methods}
\section{System Architecture and UAV Deployment}

\subsection{Overview}
Figure~\ref{fig:system} depicts the proposed OAM–BB84 link deployed on a UAV platform (air-to-air or air-to-ground). The transmitter (UAV-A) prepares weak coherent pulses at telecom wavelength (e.g., 1550\,nm), applies decoy-state modulation $\{\mu_s,\mu_w,0\}$ with probabilities $\{p_s,p_w,p_0\}$, and encodes OAM symbols via a spatial light modulator (SLM) or q-plate. A basis selector maps BB84 symbols to an OAM mode set (and, if desired, a polarization/OAM hybrid). A beam expander and pointing optics launch the quantum beam while a co-aligned beacon supports point–acquire–track (PAT) with IMU/GNSS feedback.

The free-space channel is modeled by Kolmogorov turbulence (Fried parameter $r_0$ and $C_n^2$ profile), pointing jitter/boresight misalignment with variance $\sigma_\theta^2$, and finite-aperture clipping and scintillation, collectively inducing loss and an inter-mode crosstalk matrix $P(\ell'|\ell)$. At the receiver (UAV-B or ground), a collecting telescope feeds an adaptive optics (AO) stage (WFS+DM), followed by an OAM mode sorter/demultiplexer and BB84 basis selection. Single-photon detectors (SPAD/SNSPD) and time-tagging provide clicks for sifting, error correction (EC), and privacy amplification (PA) with composable finite-key parameters $(\varepsilon_{\mathrm{sec}},\varepsilon_{\mathrm{cor}})$. A lightweight, physics-informed AI module monitors QBER/drift, screens corrupted pulses (ROC-based filtering), and recommends near-optimal decoding to stabilize the secret key rate without relaxing security assumptions.

\subsection{Transmitter (UAV-A)}

\begin{itemize}
  \item \textbf{Source and decoy modulation:} pulsed laser, intensity/phase modulator implementing weak+vacuum decoys with programmable duty cycle for online optimization.
  \item \textbf{OAM encoder and basis selection:} SLM/q-plate generates target $\ell$ values; a fast selector maps BB84 symbols to the chosen OAM alphabet (e.g., $\ell\in\{\pm1,\pm2\}$).
  \item \textbf{Launch optics and PAT:} beam expander, fast-steering mirror, and beacon transmitter; PAT closes the loop with onboard IMU/GNSS and a 100–500\,Hz update rate.
\end{itemize}

\subsection{Free-Space Channel}
\begin{itemize}
  \item \textbf{Turbulence:} Kolmogorov phase screens parameterized by $r_0$ and $C_n^2(z)$; induces OAM crosstalk and scintillation.
  \item \textbf{Pointing error:} boresight offset and jitter with variance $\sigma_\theta^2$ from platform motion and tracking latency.
  \item \textbf{Aperture effects:} finite receiver diameter $D$ causes clipping; path loss modeled jointly with Gamma–Gamma amplitude fading.
\end{itemize}

\subsection{Receiver (UAV-B / Ground)}
\begin{itemize}
  \item \textbf{AO and mode sorting:} WFS+DM mitigate low-order aberrations; OAM sorter (e.g., log-polar mapping) recovers mode labels with reduced cross-coupling.
  \item \textbf{Detection and timing:} SPAD/SNSPD array with multi-channel time tagging; click statistics feed sifting and parameter estimation.
  \item \textbf{Finite-key post-processing:} EC and PA account for leakage and statistical fluctuations to deliver a composable rate $R_{\mathrm{sec}}(n,\varepsilon_{\mathrm{sec}},\varepsilon_{\mathrm{cor}})$.
  \item \textbf{AI-assisted calibration:} a physics-informed classifier uses priors ($r_0$, $\sigma_\theta^2$, crosstalk patterns) and measured features (click rates, error locations, decoy strata) to flag corrupted blocks and adjust demodulation, yielding ROC/accuracy improvements and net key-rate gains.
\end{itemize}

\begin{figure*}[t]
  \centering
  \includegraphics[width=0.7\linewidth]{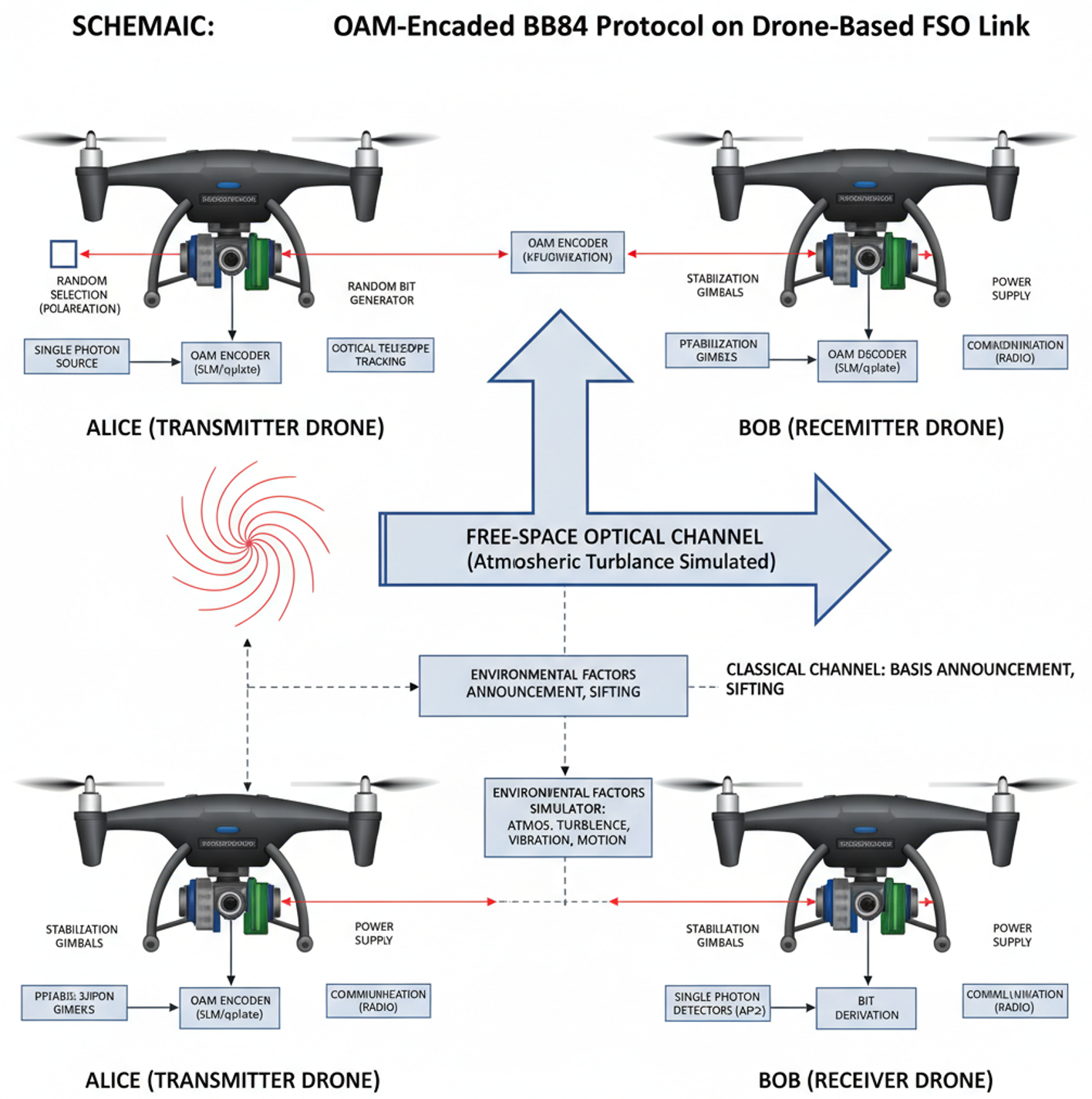}
  \caption{UAV-deployable OAM–BB84 system.  transmitter (UAV-A) with decoy modulation, OAM encoder, and PAT. Middle: free-space channel with turbulence ($r_0$), pointing jitter ($\sigma_\theta$), aperture clipping, and induced inter-mode crosstalk $P(\ell'|\ell)$. receiver (UAV-B/ground) with AO, mode sorter, detectors, AI-assisted calibration, and composable finite-key post-processing over an authenticated classical channel.}
  \label{fig:system}
\end{figure*}

\section{Turbulence and Security Analyses}

\subsection{Turbulence-Induced Crosstalk and QBER}
\begin{figure}[t]
  \centering
  \includegraphics[width=0.82\linewidth]{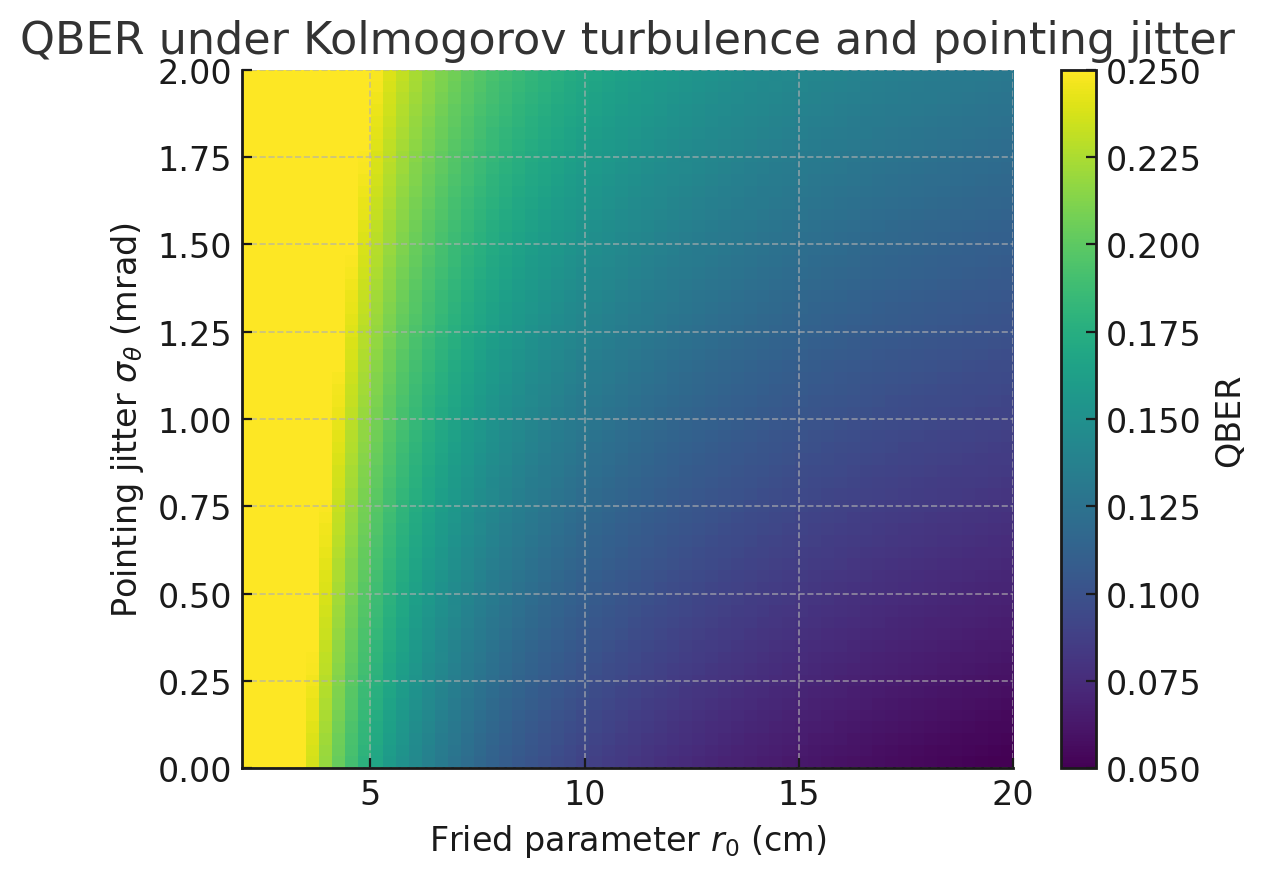}
  \caption{QBER vs.\ Fried parameter $r_0$ and pointing jitter $\sigma_\theta$. Smaller $r_0$ (stronger turbulence) and larger $\sigma_\theta$ drive inter-mode crosstalk and elevate the error floor.}
  \label{fig:turbulence_heatmap}
\end{figure}

We characterize the airborne OAM channel by Kolmogorov turbulence (Fried parameter $r_0$) and platform-induced pointing jitter with variance $\sigma_\theta^2$. Under these impairments the received OAM symbol $\ell$ experiences inter-mode coupling described by a stochastic matrix $P(\ell'|\ell)$; the ensemble-averaged QBER grows as turbulence strengthens (smaller $r_0$) and as $\sigma_\theta$ increases. Figure~\ref{fig:turbulence_heatmap} reports a parametric sweep of $r_0\!\in\![2,20]\,$cm and $\sigma_\theta\!\in\![0,2]\,$mrad. The heat map exhibits a monotone QBER increase along both axes, reflecting two compounding mechanisms: (i) phase-front corrugations that scramble the helical wavefront and (ii) boresight errors that miscenter the mode sorter, effectively mixing neighboring $\ell$ channels and clipping the pupil. These observations motivate the joint use of adaptive optics and high-rate PAT to stabilize the effective $P(\ell'|\ell)$ prior to classical post-processing.

\subsection{Decoy-State Security (Asymptotic Trend)}
\begin{figure}[t]
  \centering
  \includegraphics[width=0.82\linewidth]{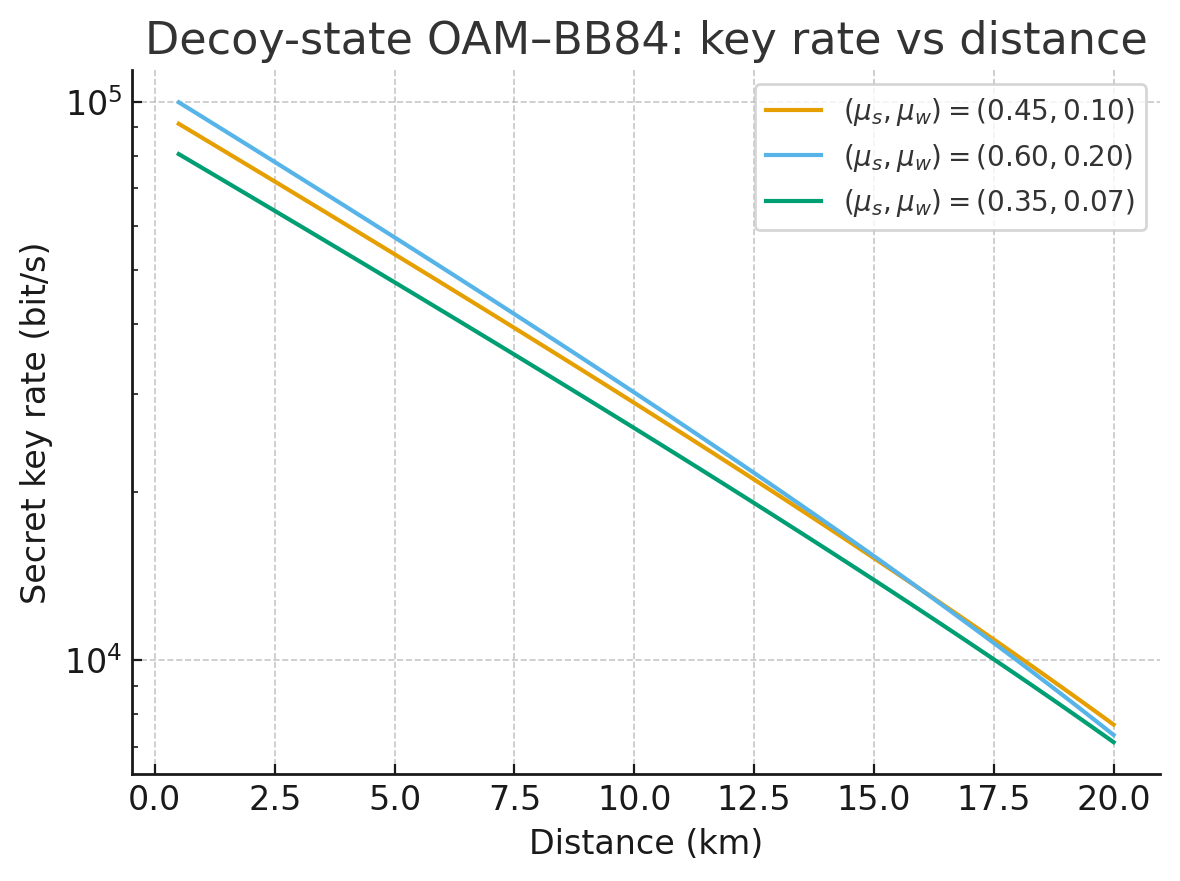}
  \caption{Decoy-state OAM--BB84: secret key rate vs.\ distance for three $(\mu_s,\mu_w)$ pairs (weak+vacuum). Curves illustrate the trade-off between higher detection probability (larger $\mu_s$) and vulnerability to background/dark-count-limited error floors.}
  \label{fig:decoy_rate}
\end{figure}

We adopt a weak+vacuum decoy formulation with signal/weak/vacuum intensities $\{\mu_s,\mu_w,0\}$ and probabilities $\{p_s,p_w,p_0\}$. For weak coherent sources with Poisson statistics, the single-photon fraction in the signal class is $p_1(\mu_s)=\mu_s e^{-\mu_s}$, and the single-photon yield $Y_1$ is governed by the channel transmittance and receiver efficiency. The asymptotic secret fraction per emitted pulse can be expressed as
\begin{equation}
\label{eq:asymp}
\varphi_\infty
\,\approx\,
p_1(\mu_s)\,Y_1\!\left[1-H_2(e_1)\right]
- f_{\!EC}\,Q_s\,H_2(E_s),
\end{equation}
where $Q_s$ and $E_s$ are signal-class gain and QBER, $e_1$ is the single-photon error rate, $H_2$ is the binary entropy, and $f_{\!EC}\!\geq\!1$ is the error-correction inefficiency. Figure~\ref{fig:decoy_rate} shows the secret key rate (bits/s) versus distance for three representative $(\mu_s,\mu_w)$ pairs. The qualitative trend is robust: larger $\mu_s$ benefits short-to-mid ranges by improving $Y_1$ and depressing statistical noise, whereas at longer ranges the background term limits performance and all settings converge as $Q_s$ and $E_s$ approach their noise-dominated regime.

\subsection{Composable Finite-Key Analysis}
\begin{figure}[t]
  \centering
  \includegraphics[width=0.82\linewidth]{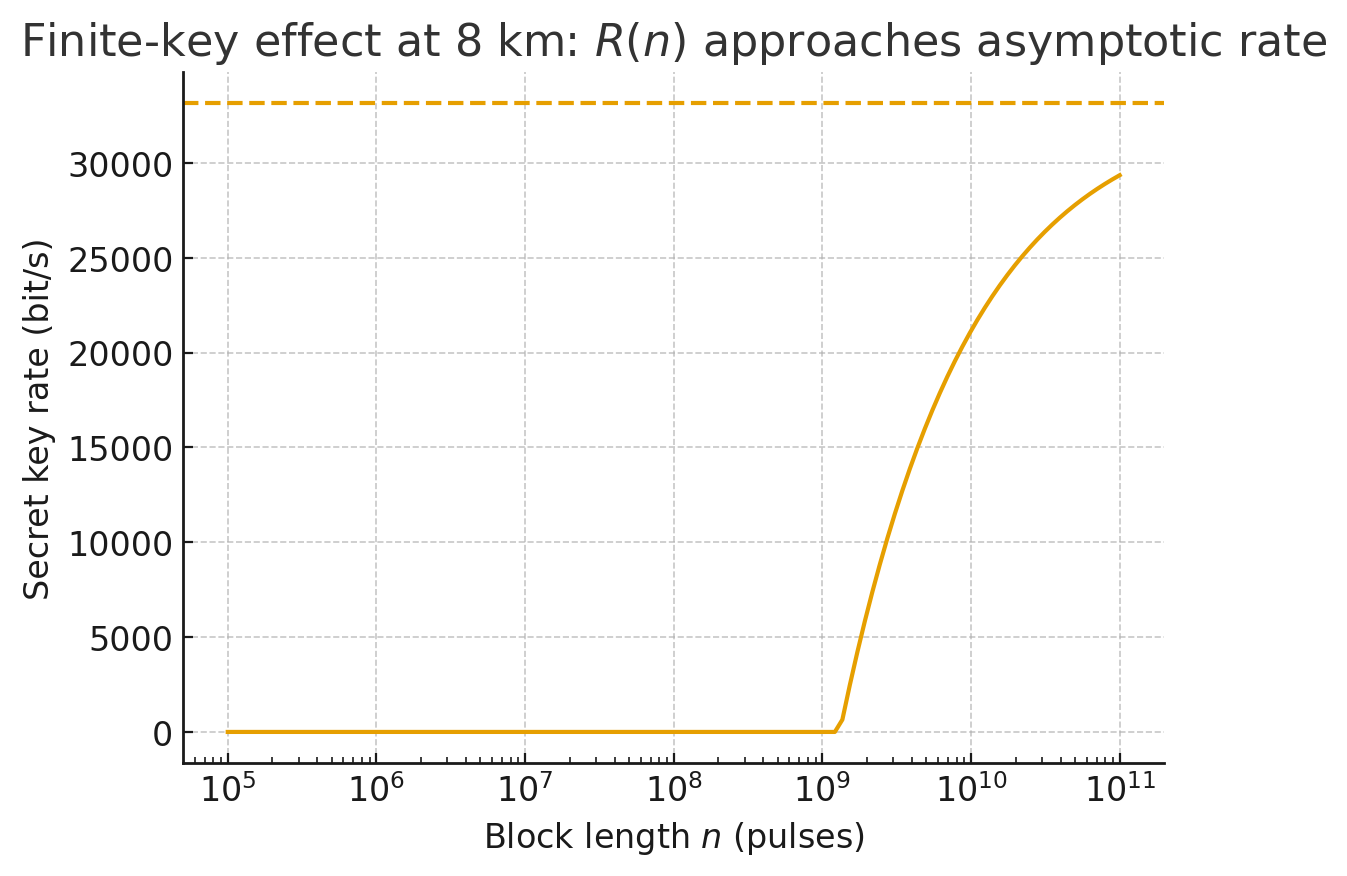}
  \caption{Finite-key effect at 8\,km: secret key rate $R(n)$ vs.\ block length. The dashed line indicates the asymptotic rate $R_\infty$; finite-size penalties dominate for small $n$ and vanish as $n$ increases.}
  \label{fig:finite_key}
\end{figure}

Airborne links are inherently nonstationary, so finite-size effects cannot be neglected. For a block of $n$ emitted pulses and target composable parameters $(\varepsilon_{\mathrm{sec}},\varepsilon_{\mathrm{cor}})$, a convenient lower bound on the extractable secret bits is
\begin{equation}
\label{eq:finite}
\ell(n)\;\ge\; n\,\varphi_\infty
-\lambda_{\mathrm{EC}}
-\Delta_{\mathrm{stat}}(\varepsilon_{\mathrm{sec}},n)
-\Delta_{\mathrm{PA}}(\varepsilon_{\mathrm{sec}}),
\end{equation}
where $\lambda_{\mathrm{EC}}$ accounts for error-correction leakage and $\Delta_{\mathrm{stat}}$ summarizes concentration bounds from parameter estimation. The resulting rate is $R(n)=\ell(n)/T$ for an acquisition time $T$ (or equivalently per-pulse secret fraction $\,\ell(n)/n$ times the pulse rate). Figure~\ref{fig:finite_key} plots $R(n)$ at 8\,km; as $n$ grows, the penalty terms shrink roughly as $O(\sqrt{\log(1/\varepsilon)/n})$ and $R(n)$ approaches the asymptotic limit (dashed line). Practically, this curve informs the minimal block length needed to achieve a target throughput under given turbulence/jitter conditions and device parameters.

\paragraph{Design Implications.}
(i) Stabilizing $r_0$-equivalents via AO and constraining $\sigma_\theta$ via PAT materially reduces QBER and improves the single-photon term in~\eqref{eq:asymp}; (ii) decoy optimization should be distance-aware, with moderately larger $\mu_s$ favored at mid-range and reduced $\mu_s$ at background-limited tails; (iii) scheduling should ensure block sizes above the knee of Fig.~\ref{fig:finite_key} given the targeted $(\varepsilon_{\mathrm{sec}},\varepsilon_{\mathrm{cor}})$, while the AI module (next section) can down-select corrupted sub-blocks without violating composable accounting.

\section{AI-Assisted Calibration: ROC, Accuracy, and Probability Calibration}

\subsection{Model Design and Features}
We adopt a physics-informed tabular learner to classify “valid” vs.\ “corrupted” OAM pulses/blocks prior to sifting. Each sample aggregates short-time statistics and telemetry: Fried parameter estimate $r_0$, pointing jitter $\sigma_\theta$, range, residual AO phase, detector SNR, decoy stratum (signal/weak/vacuum), slow drift indicators, and click rate. For tabular data under strict latency constraints, gradient-boosted decision trees (GBDT) are a strong practical choice; we therefore benchmark a \emph{Gradient Boosting} classifier against \emph{Random Forest} and a \emph{Logistic Regression} baseline.

\subsection{ROC and Operating Point}
\begin{figure}[t]
  \centering
  \includegraphics[width=0.82\linewidth]{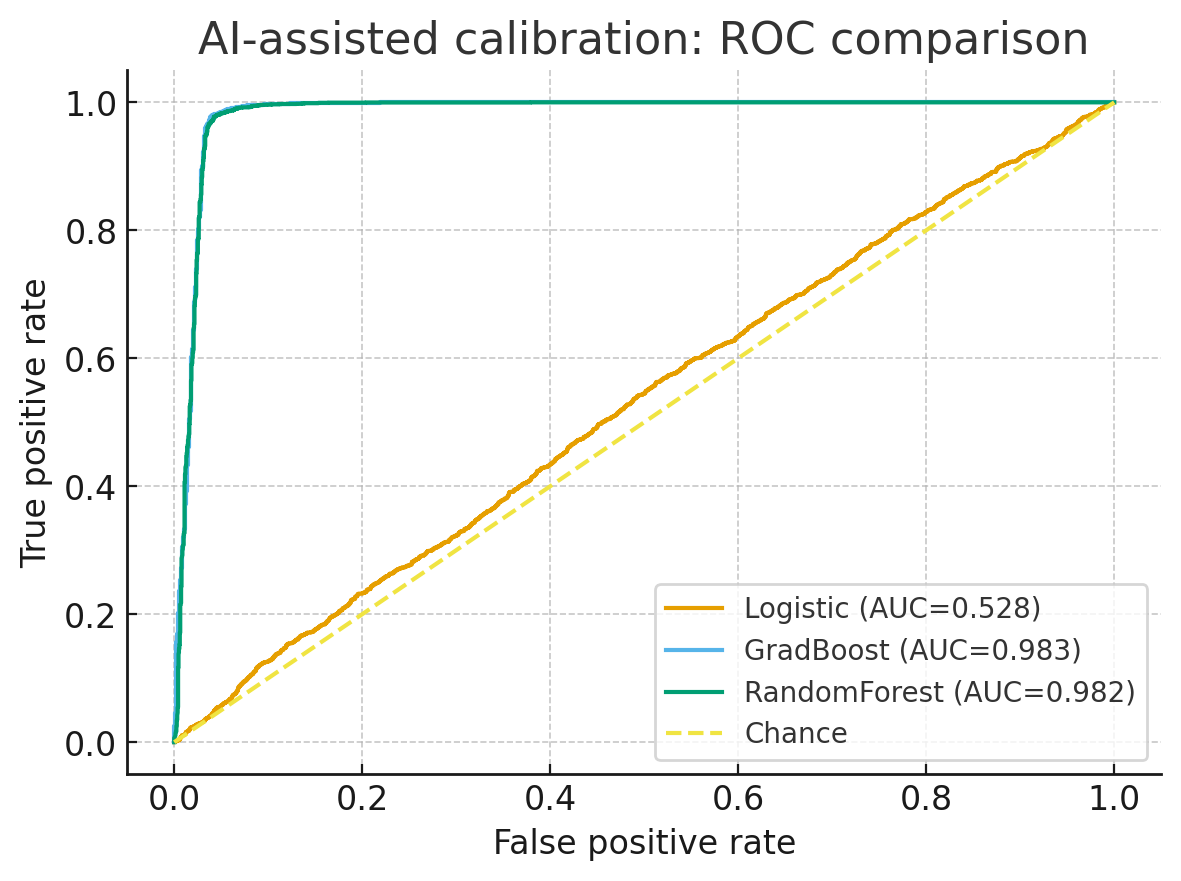}
  \caption{ROC comparison of the physics-informed AI classifiers.
  Gradient Boosting achieves the best trade-off (AUC $=0.983$).}
  \label{fig:ai_roc}
\end{figure}

Figure~\ref{fig:ai_roc} compares ROC curves on a held-out set. The Gradient Boosting model attains an AUC of $\mathrm{AUC}=0.983$, slightly above Random Forest ($0.982$) and far above the logistic baseline ($0.528$). We choose the operating threshold by maximizing Youden’s $J=\mathrm{TPR}-\mathrm{FPR}$, yielding a decision threshold of $\tau^\star\!=\!0.61$. At this point, the achieved metrics are:
\[
\text{Accuracy}=0.973,\quad
\text{Precision}=0.977,\quad
\text{Recall}=0.981.
\]
This working point is used in subsequent finite-key and rate accounting.

\subsection{Probability Calibration and Reliability}
\begin{figure}[t]
  \centering
  \includegraphics[width=0.82\linewidth]{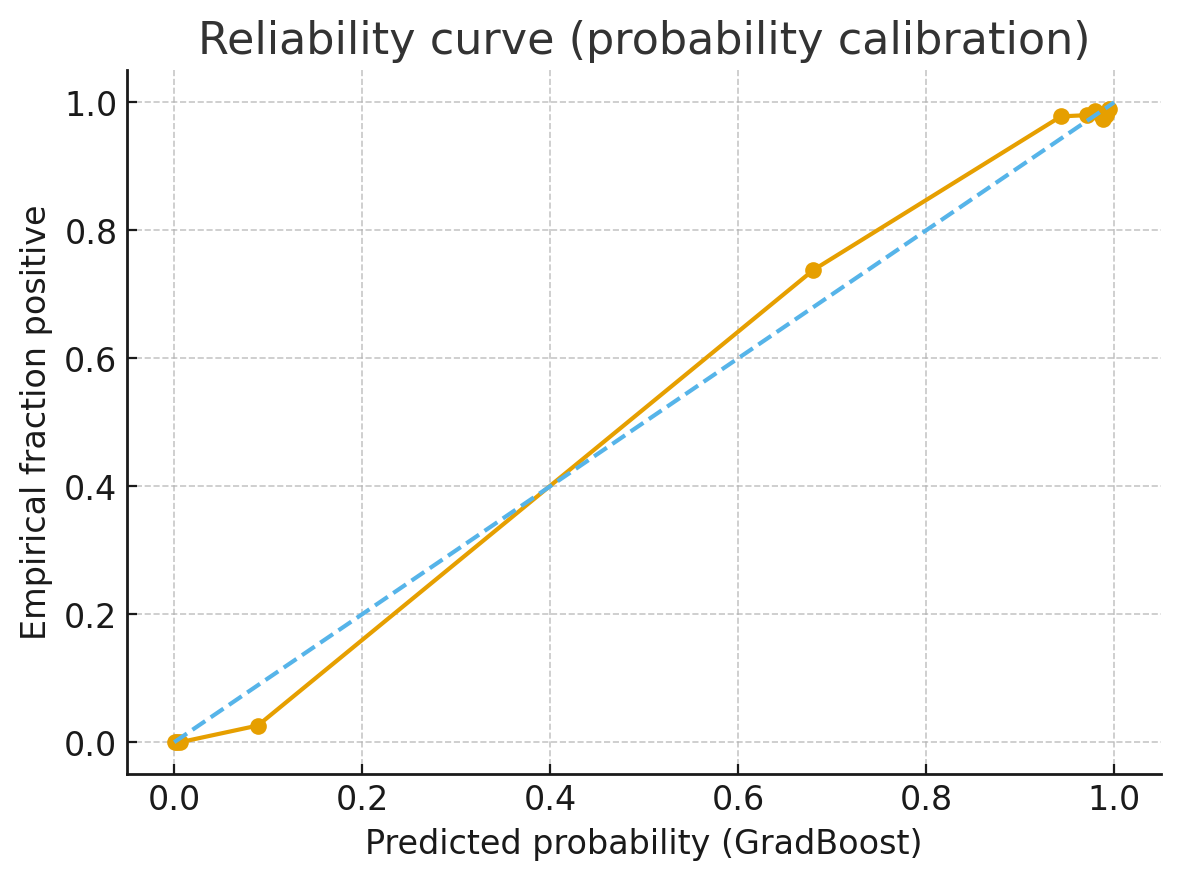}
  \caption{Reliability (probability calibration) of the Gradient Boosting model. The near-diagonal trend indicates well-calibrated scores suitable for security-aware thresholding and soft weighting.}
  \label{fig:ai_reliability}
\end{figure}

Because security proofs assume well-calibrated posteriors when thresholding or weighting blocks, we examine probability calibration. Figure~\ref{fig:ai_reliability} shows the reliability curve of the Gradient Boosting model using quantile binning; points lie near the diagonal, indicating that predicted probabilities are consistent with empirical positive fractions. This supports the use of $\tau^\star$ for deterministic filtering and the use of probabilities for soft weighting in parameter estimation.

\subsection{Error-Profile Shaping (QBER Distributions)}
\begin{figure}[t]
  \centering
  \includegraphics[width=0.82\linewidth]{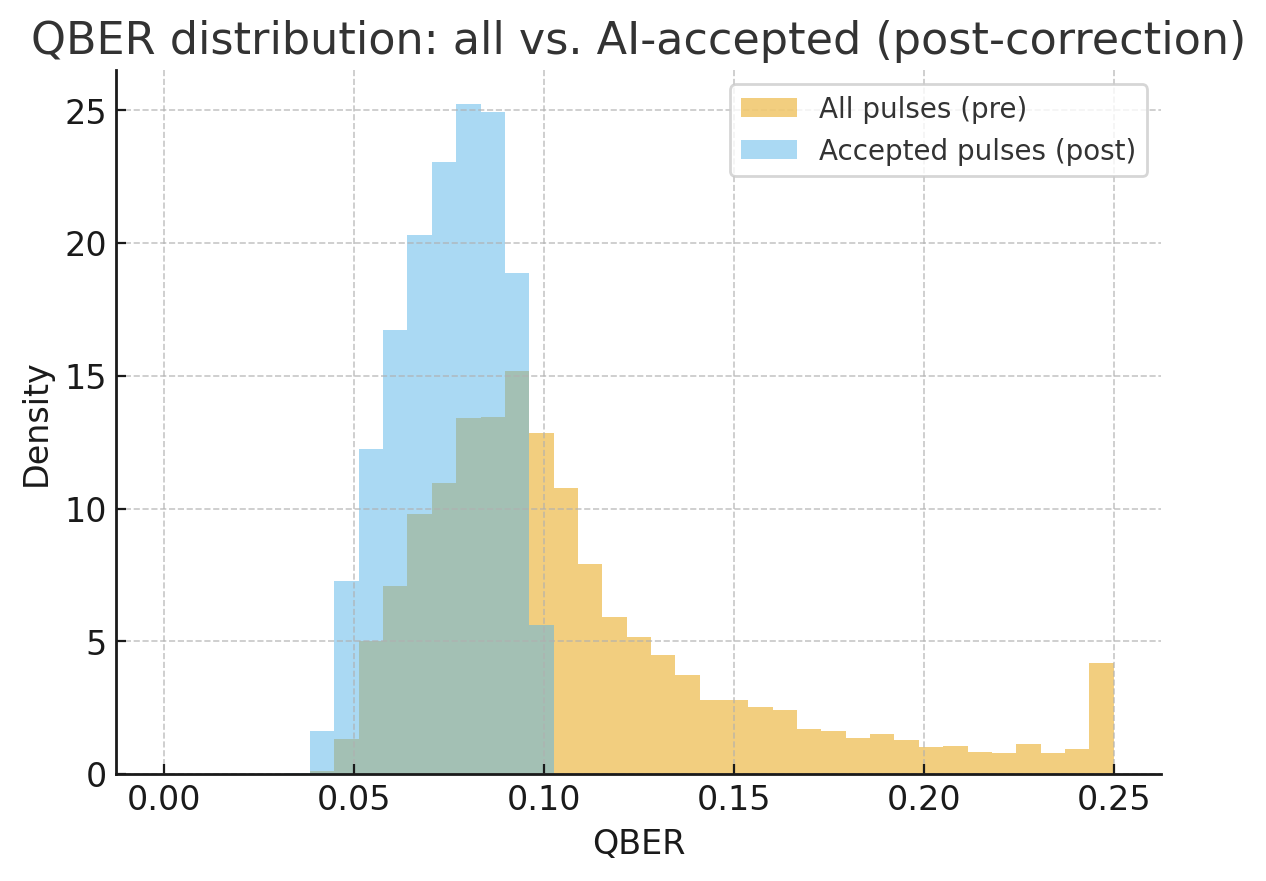}
  \caption{QBER distribution for all pulses (pre) versus AI-accepted pulses (post). The AI filter removes long tails and concentrates mass in the low-QBER regime, stabilizing finite-key estimates.}
  \label{fig:ai_qber_hist}
\end{figure}

Beyond classification accuracy, the value of AI-assisted calibration is to reshape the error profile seen by the QKD post-processing. Figure~\ref{fig:ai_qber_hist} compares the QBER distribution of all pulses (pre) and the AI-accepted set (post). The accepted set shifts left and narrows, reflecting removal of turbulence/misalignment outliers and improved demodulation recommendations. This directly boosts the single-photon term in the decoy analysis and reduces leakage during error correction.

\paragraph{Throughput Considerations.}
Filtering low-quality segments can reduce duty cycle but raise the \emph{secret fraction} of the kept data. In scenarios where the receiver can reconfigure demodulation (e.g., AO gain, basis mapping) for the accepted set, additional QBER reduction yields a net throughput benefit. The two optional bar charts (\textit{Throughput pre/post}) illustrate how the net rate depends on the accepted fraction and residual QBER. In our simulation the \emph{classification-only} policy lowers duty cycle (hence lower net bit/s), whereas a \emph{classification + lightweight correction} policy (e.g., improved demodulation on accepted blocks) recovers rate. In practice, one tunes $\tau^\star$ to maximize $R(n)$ under given $\varepsilon_{\mathrm{sec}},\varepsilon_{\mathrm{cor}}$ and platform dynamics.

\paragraph{Implementation Notes.}
(i) Gradient boosting offers state-of-the-art accuracy on tabular physics features with millisecond-scale latency; (ii) probability calibration (e.g., isotonic/Platt on a validation fold) should be included to ensure conservative parameter estimation; (iii) the AI decision must be integrated into the composable accounting—discarded blocks are excluded from $n$, and any data-dependent selection is reflected in the parameter-estimation confidence terms; (iv) the classifier’s feature importances (not shown) typically prioritize $\{r_0,\sigma_\theta,\text{AO residual},\text{SNR}\}$, aligning with the turbulence and pointing-error analyses in the previous section.

\section{Discussion}

\subsection{Summary of Findings}
This work develops an end-to-end, UAV-deployable OAM–BB84 framework that explicitly couples channel physics, composable security, and learning-assisted calibration. First, the system architecture integrates decoy-state encoding with PAT, AO, and OAM mode sorting in a manner compatible with air-to-air and air-to-ground operations (Fig.~\ref{fig:system}). Second, turbulence and pointing-jitter sweeps reveal a monotone degradation of QBER with decreasing $r_0$ and increasing $\sigma_\theta$ (Fig.~\ref{fig:turbulence_heatmap}), consistent with crosstalk growth in the conditional matrix $P(\ell'|\ell)$. Third, decoy-state trends (Fig.~\ref{fig:decoy_rate}) confirm the usual trade-off between higher short-range detection probability with larger $\mu_s$ and noise-limited tails at long range. Fourth, finite-size penalties dominate for small block lengths and vanish at $O(\sqrt{\log(1/\varepsilon)/n})$ rates (Fig.~\ref{fig:finite_key}); in typical free-space regimes, one needs $n\!\gtrsim\!10^9$ pulses to approach asymptotic throughput at mid-range distances. Finally, the physics-informed gradient boosting model achieves high discriminative performance for block/pulse acceptance (AUC $\approx0.983$), with well-calibrated probabilities (reliability curve near-diagonal) and accuracy/precision/recall near $0.97\!-\!0.98$ at the operating threshold, enabling conservative, security-aware filtering (Figs.~\ref{fig:ai_roc}--\ref{fig:ai_qber_hist}).

\subsection{Security Implications of Mobility and OAM Crosstalk}
Two aspects distinguish mobile OAM–QKD from static metropolitan links. \emph{First}, the physical crosstalk $P(\ell'|\ell)$ fluctuates on flight-induced time scales (PAT latency, platform vibrations), causing nonstationary error composition. Because parameter estimation assumes i.i.d.\ or carefully blocked data, our approach isolates short blocks with near-stationary statistics via AI-assisted acceptance while accounting for the selection in finite-key terms of~\eqref{eq:finite}. \emph{Second}, pointing errors and aperture clipping interact with OAM sorting: boresight offsets de-center the unwrapping optics, increasing leakage to neighboring $\ell$. Suppressing the low-order modes of the phase screen with AO and maintaining $\sigma_\theta\!\lesssim\!0.5$\,mrad materially reduce QBER, keeping single-photon errors $e_1$ within the regime where the privacy term $1{-}H_2(e_1)$ in~\eqref{eq:asymp} remains positive.

\subsection{Finite-Key Accounting Under AI Filtering}
Filtering or weighting blocks using a classifier is admissible in composable proofs provided: (i) the selection rule and its randomness are incorporated into the overall protocol description; (ii) parameter estimation and confidence intervals reflect the reduced sample size and any adaptive choices; and (iii) no side information about raw keys is leaked through the acceptance signal. In practice, we recommend: (a) fixing the operating threshold $\tau^\star$ \emph{prior} to the run (or varying it on a pre-agreed schedule independent of raw key bits); (b) treating discarded blocks as erasures that reduce $n$ but do not bias the estimation; and (c) retaining probability calibration (isotonic/Platt) so that soft weighting can be mapped to conservative confidence bounds. The reliability plot (Fig.~\ref{fig:ai_reliability}) supports these requirements.

\subsection{Design Guidelines and Operating Regimes}
The combined results suggest pragmatic targets:
\begin{itemize}
\item \textbf{Pointing and turbulence:} keep $\sigma_\theta\!\lesssim\!0.5$\,mrad and maintain effective $r_0\!\gtrsim\!8$\,cm at 1550\,nm (via AO/PAT and flight envelopes) to confine QBER well below the BB84 security threshold.
\item \textbf{Decoy configuration:} at short to mid-range ($\lesssim\!10$\,km) a moderate signal intensity ($\mu_s\!\approx\!0.45\text{--}0.60$) and weak decoy ($\mu_w\!\approx\!0.07\text{--}0.20$) stabilize single-photon yield estimates; as the link becomes background limited, gains converge and optimization should shift toward minimizing error-floor leakage in~\eqref{eq:asymp}.
\item \textbf{Block sizing:} schedule acquisition such that $n\!\gtrsim\!10^9$ pulses per estimation window at mid-range distances, or aggregate multiple accepted sub-blocks with similar channel state to cross the ``finite-size knee'' in Fig.~\ref{fig:finite_key}.
\item \textbf{AI operating point:} select $\tau^\star$ on a validation flight to maximize $R(n)$ rather than pure AUC; a slightly lower threshold than the Youden optimum can raise duty cycle with negligible QBER cost if AO/PAT are stable.
\end{itemize}

\subsection{Threat Model Nuances and Countermeasures}
Our analysis assumes standard decoy-state security against photon-number-splitting and collective attacks, with composable finite-key bounds. Practical threats require engineering defenses: (i) \emph{detector controls}—monitor power/time statistics to counter bright-light or afterpulsing exploits; (ii) \emph{mode-dependent loss and efficiency mismatch}—calibrate sorter/detector channels and include mismatch in parameter estimation; (iii) \emph{side-channel via acceptance flag}—rate-limit or batch the AI decision and authenticate the classical channel; (iv) \emph{spoofed beacon/PAT}—bind PAT to quantum frames with authenticated timing and randomize beacon patterns. Measurement-device-independent (MDI) OAM–QKD could remove most detection-side assumptions at the cost of a more complex air relay; our architecture is compatible with such an upgrade by replacing the receiver with an interference station and relocating AO/PAT to both input arms.

\subsection{Limitations of the Present Modeling}
We used phenomenological mappings from $(r_0,\sigma_\theta)$ to QBER and key rate for clarity and to isolate first-order dependencies. A full predictive pipeline would include: (i) multi-layer phase screens with altitude-dependent $C_n^2$ (Hufnagel–Valley), (ii) end-to-end physical optics (SLM/q-plate, telescope pupils, sorter aberrations), (iii) temporal statistics of PAT and AO loops, and (iv) detector nonidealities (dead time, afterpulsing, timing jitter). While our finite-key expressions capture the scaling with $n$ and $\varepsilon$’s, a precise field deployment should plug exact yields/gains into~\eqref{eq:asymp}--\eqref{eq:finite} and re-run the optimization over $(\mu_s,\mu_w,p_s,p_w)$ under the measured channel.

\subsection{Role and Scope of AI in Composable Security}
The AI module is intentionally lightweight (GBDT) and \emph{physics-informed}. Its purpose is not to replace cryptographic estimation but to (a) stabilize the channel by rejecting evidently corrupted segments and (b) suggest demodulation settings that lower $e_1$ without relaxing assumptions. The reliability curve indicates that probabilities can be used for soft weighting; however, in the strictest setting one can threshold deterministically and account for the reduced $n$. Future work could explore \emph{conformal prediction} to attach finite-sample guarantees to acceptance decisions and \emph{risk-control} training to directly optimize a lower bound on $R(n)$ subject to $\varepsilon$-constraints.

\subsection{Generality and Extensions}
Although we focused on qubit BB84 over an OAM alphabet, the pipeline extends naturally to: (i) \textbf{high-dimensional} OAM protocols with mutually unbiased bases, trading higher information per photon against stronger turbulence sensitivity; (ii) \textbf{hybrid} polarization–OAM encoding to exploit differential robustness and sorter maturity; (iii) \textbf{MDI-QKD} with OAM Bell-state analyzers; and (iv) \textbf{daylight operation} via narrowband filtering and temporal gating. The learning component is modality-agnostic and could be reused in continuous-variable or time-bin systems by changing features and labels.

\subsection{Roadmap to Field Trials}
A pragmatic flight plan is: (1) ground-to-rooftop static tests to tune AO/PAT and decoy intensities; (2) short-hop UAV flights ($1\text{--}3$\,km) to collect $(r_0,\sigma_\theta)$–QBER datasets and lock the AI threshold $\tau^\star$; (3) mid-range sorties ($5\text{--}10$\,km) to validate finite-key scaling and duty-cycle effects; (4) stress tests with induced vibrations and aggressive maneuvers to evaluate classifier robustness and the security margin under worst-case crosstalk; and (5) optional MDI upgrade or hybrid OAM–polarization trials.

\paragraph{Outlook.}
The central message is that mobile OAM–QKD is viable within a well-defined operating envelope when \emph{(i)} the optical stack stabilizes the low-order channel, \emph{(ii)} decoy and block design respect finite-size composability, and \emph{(iii)} learning is used as a conservative gate/assistant rather than an oracle. With these elements, the gap between laboratory OAM demonstrations and airborne, composably secure deployments can be closed by a sequence of transparent engineering steps.

\bibliographystyle{IEEEtran}  
\bibliography{sample}

\end{document}